\documentclass[conference]{IEEEtran}
\IEEEoverridecommandlockouts
\usepackage{cite}
\usepackage{amsmath,amssymb,amsfonts}
\usepackage{algorithmic}
\usepackage{graphicx}
\usepackage{textcomp}
\usepackage{xcolor}
\def\BibTeX{{\rm B\kern-.05em{\sc i\kern-.025em b}\kern-.08em
    T\kern-.1667em\lower.7ex\hbox{E}\kern-.125emX}}

\begin{document}

\title{EEG-Transformer: Self-attention from \\Transformer Architecture for Decoding EEG of \\Imagined Speech\\
\thanks{This work was partly supported by Institute for Information \& Communications Technology Planning \& Evaluation (IITP) grant funded by the Korea government (MSIT) (No. 2017-0-00451, Development of BCI based Brain and Cognitive Computing Technology for Recognizing User’s Intentions using Deep Learning; No. 2015-0-00185, Development of Intelligent Pattern Recognition Softwares for Ambulatory Brain Computer Interface; No.2021-0-02068, Artificial Intelligence Innovation Hub).}
}

\author{
\IEEEauthorblockN{Young-Eun Lee}
\IEEEauthorblockA{\textit{Dept. Brain and Cognitive Engineering} \\
\textit{Korea University}\\
Seoul, Republic of Korea \\
ye\_lee@korea.ac.kr}
\and
\IEEEauthorblockN{Seo-Hyun Lee}
\IEEEauthorblockA{\textit{Dept. Brain and Cognitive Engineering} \\
\textit{Korea University}\\
Seoul, Republic of Korea \\
seohyunlee@korea.ac.kr}

}

\maketitle

\begin{abstract}

Transformers are groundbreaking architectures that have changed a flow of deep learning, and many high-performance models are developing based on transformer architectures. Transformers implemented only with attention with encoder-decoder structure following seq2seq without using RNN, but had better performance than RNN. 
Herein, we investigate the decoding technique for electroencephalography (EEG) composed of self-attention module from transformer architecture  during imagined speech and overt speech. We performed classification of nine subjects using convolutional neural network based on EEGNet that captures temporal-spectral-spatial features from EEG of imagined speech and overt speech. Furthermore, we applied the self-attention module to decoding EEG to improve the performance and lower the number of parameters. Our results demonstrate the possibility of decoding brain activities of imagined speech and overt speech using attention modules. Also, only single channel EEG or ear-EEG can be used to decode the imagined speech for practical BCIs.
\end{abstract}

\begin{small}
\end{small}

\section{Introduction}
% BCI
Brain-computer interfaces (BCIs) are one of the most important consideration for communication systems in real life. Many researchers have studied BCI to recognize human cognitive state or intention based on brain signals such as electroencephalography (EEG) to recognize the crucial features from the brain activity. \cite{zhang2017hybrid,jeong2020brain,zhang2019strength}.
To enhance the performance of decoding EEG signals, preprocessing technology is also important to get a high quality signals with higher accuracy of decoding and lower signal-to-noise ratio \cite{castermans2011optimizing,lee2020real,kwak2015lower,lee2020reconstructing}. Moreover, the decoding technologies including feature extraction and classification have improved significantly in recent years \cite{kwak2017convolutional,nordin2018dual,lee2017network,kwon2019subject,lee2020decoding}.

% imagined speech
Recognizing brain activities during speech or imagined speech has recently attracted a lot of attention and is developing \cite{huth2016natural,schoffelen2017frequency}. In particular,imagined speech is evaluated as an advanced technology for brain signal-based communication systems \cite{wolpaw2002brain, lee2020neural, lee2018high}. Imagined speech refers to the internal pronunciation of speech only by imagination without auditory output or pronunciation \cite{schultz2017biosignal}. 
Recent studies have shown some features and potentials of imagined speech decoding \cite{lee2020neural, nguyen2017inferring}, but fundamental neural properties and their practical use remain to be investigated. Therefore, research on the decoding of imagined speech requires the development of brain signal decoding techniques for more accurate and practical BCI \cite{lee2019eeg, anumanchipalli2019speech}.

% Deep learning for EEG (EEG에 딥러닝이 사용된다)
Several deep learning techniques have been published to decode EEG brain signals, which are architectural designs that considers the characteristics of brain signal characteristics \cite{schirrmeister2017deep, lawhern2018eegnet}. It was often used to decode human intention using motor imagery or event-related potential, and have shown superior performance than the conventional machine learning methods such as linear discriminant analysis and support vector machine \cite{suk2012novel,kwon2019subject, gao2020classification}. Recently, there are several attempts to find optimal features of EEG by deep neural networks based on the three main features of EEG, temporal, spectral, and spatial features \cite{waytowich2018compact, bhatti2019soft}. In addition, EEG-based speaker identification studies also have actively applied machine learning or  deep learning techniques \cite{moctezuma2019subjects, dash2019spatial}. Deep learning may be effective in capturing prominent features from brain signals to verify individual characteristics.

\begin{figure*}[t]
\centering
    \includegraphics[width=\textwidth]{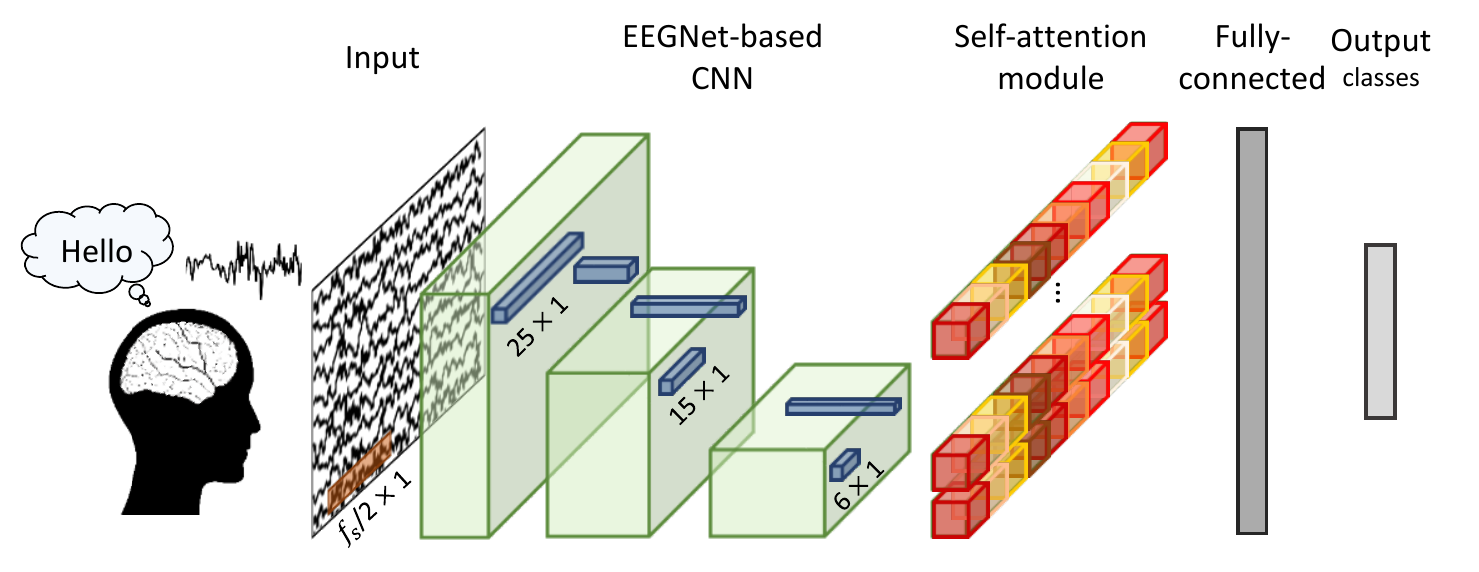}
    \caption{Total frameworks in this study. We split an image into fixed-size patches, linearly embed each of them,
add position embeddings, and feed the resulting sequence of vectors to a standard Transformer encoder. In order to perform classification, we use the standard approach of adding an extra learnable classification token to the sequence.}
    \label{fig1}
\end{figure*}

% transformer
Transformer\cite{vaswani2017attention} is a model from Google's 2017 paper ``Attention is all you need," and is implemented only with attention, while following the encoder-decoder, the existing structure of seq2seq. This model does not use RNN, and even though the encoder-decoder structure is designed, it also has better performance than RNN. This is the basis for famous language models such as GPT-3 and DALL-E, and tools such as the Hugging Face Transformers library have made it easy for machine learning engineers to solve a wide range of NLP tasks and have since promoted numerous innovations in NLP and other fields \cite{zhao2020exploring, zhang2019self, parmar2018image, dosovitskiy2020image}.
Transformer's attention was created to overcome the limitations of RNN, which was slow in computation due to difficulties in parallel processing. Transformers do not need to process data sequentially like RNNs. In addition, this processing method is possible because it allows much more parallelization than RNN.

% practical BCI
Recently, there have been several attempts to commercialize BCI technology\cite{fiedler2017single, debener2015unobtrusive}. For example, portable and non-hair EEGs were frequently investigated to improve the applicability of BCI in real life, and endogenous paradigms such as motor imagery and imagined speech are used rather than exogenous paradigm such as event-related potential and steady-state visual evoked potential, which needs external devices to give stimuli\cite{lee2021mobile}. In particular, the ear-EEG composed of electrodes disposed inside or around the ear has many advantages over the existing scalp-EEG in terms of stability and portability. In addition, since the Broca-Wernicke region, which is mainly analyzed during overt speech or imagined speech, is distributed close to the left ear channels, there is a possibility that only a small number of channels can be used to recognize the user's intention \cite{huth2016natural,lee2020neural}.

\section{Materials and Methods}

\subsection{Data Description}
The experimental protocol followed the previous works \cite{lee2020neural,lee2019eeg}. Nine subjects (three males; age 25.00 ± 2.96) participated in the study. The study was approved by the Korea University Institutional Review Board [KUIRB-2019-0143-01] and was conducted in accordance with the Declaration of Helsinki. Informed consent was obtained from all subjects.  

We recorded EEG signals from scalp during overt speech and imagined speech. After recording two seconds of resting state, two more seconds of voice audio for each word/phrase were provided, followed by consecutive trials of imagined speech or overt speech \cite{lee2020neural, lee2019towards}. During the experiment in which each block was repeated the imagined or overt speech four times, only the first trial of each block was used to match the number of trials with different experimental conditions. Each participant conducted a random experiment 25 times for every 12 words, and a total of 300 trials for each condition.
There are 13 classification outputs, consisting of 12 words (ambulance, clock, hello, help me, light, pain, stop, thank you, toilet, TV, water, and yes) and resting state.

\subsection{EEG Preprocessing}
The EEG signal was down-sampled at 250 Hz and divided into 2 seconds from the start of each trial. The preprocessing of EEG signal was performed with a 5th Butterworth filter in the high-gamma band of 30--120 Hz, and baseline was corrected by subtracting the average of 500 ms before the start of each trial. We selected the channels located in the Broca and Wernicke’s areas (AF3, F3, F5, FC3, FC5, T7, C5, TP7, CP5, and P5). For removing the artifacts of EOG and EMG from muscle activity around mouth, we conducted artifact removal methods using independent component analysis with references from EOG and EMG. 
All data processing procedures were performed in Python and Matlab using OpenBMI Toolbox \cite{leeMH2019eeg}, BBCI Toolbox \cite{krepki2007berlin}, and EEGLAB \cite{Delorme2004EEGLAB}.

\begin{figure}[t]
\centering
    \includegraphics[width=\columnwidth]{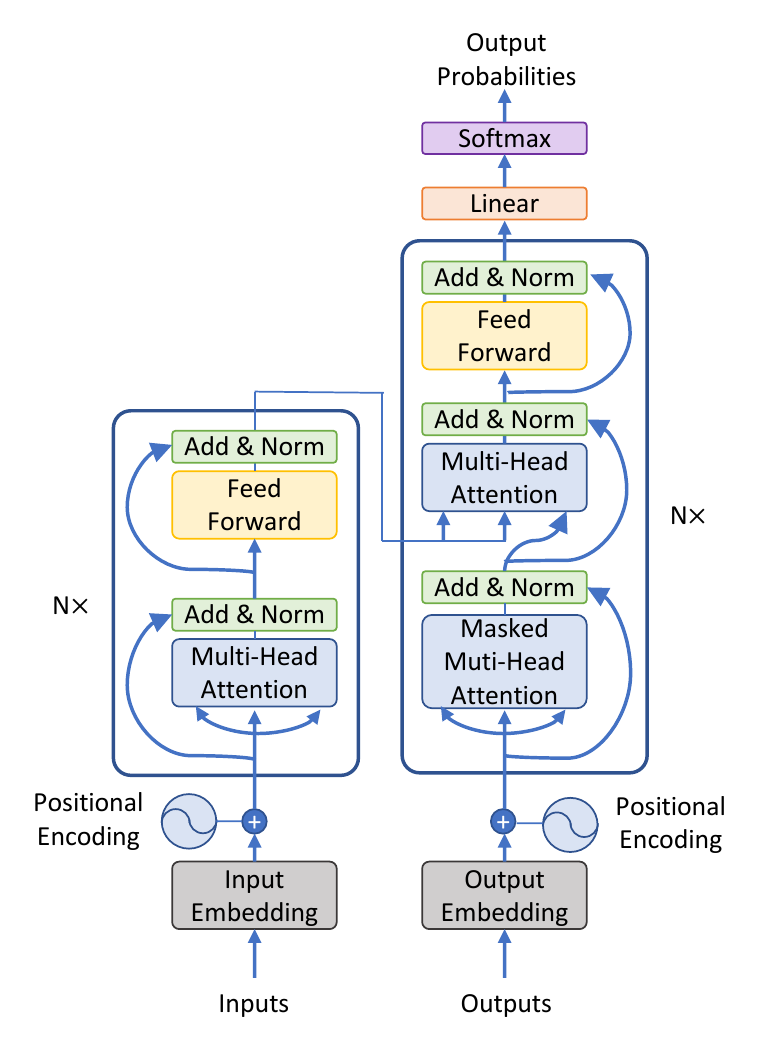}
    \caption{Transformer architecture. Self-attention and the feed-forward networks are followed by dropout and bypassed by a residual connection with subsequent layer normalization. This figure is inspired from the Transformer paper \cite{vaswani2017attention}.}
    \label{fig2}
\end{figure}

\subsection{Architecture}
The proposed classification framework consists of convolution layers and separable convolution layers for extracting time-spectral-spatial information, as shown in Fig. \ref{fig1}. Given the input as raw signals (C $\times$ T), classification output is set to 13 classes. The kernel size of the first layer is set in relation to the sampling frequency of the data for performing a temporal convolution that imitates the band-pass filter \cite{waytowich2018compact}. Since support vector machine (SVM) classifier has been reported to be robust in decoding the imagined speech \cite{lee2020neural,nguyen2017inferring}, we used the squared hinge loss for training that functions similar to the margin loss of SVM.
The evaluation was conducted through 5-fold cross-validation and 1000 epoch training for each condition. The probability of chance level of this experiment was 11.11\% because the number of samples of each subject were the same for each state condition.

The self-attention module was shown is Fig. \ref{fig1} and Fig. \ref{fig2} as well. The attention module serves for mapping a set of query and key-value pairs, where the output is calculated as the weighted sum of the values, where the weights assigned to each value are calculated by the compatibility function of the query and its key. The multi-head attention can jointly focus on information in different representation subspaces at different positions, resulting in an average calculation with one attention.

\subsection{Statistical Analysis}
Statistical analysis were performed to verify the results of classification. Kruskal-Wallis non-parametric one-way analysis of variance (ANOVA) were performed to compare the classification performance of imagined speech and overt speech. Post-hoc analysis was conducted with non-parametric permutation-based t-test. The Kruskal-Wallis test was also performed on classification performance using a single channel EEG to estimate the significance of the selected channel. In addition, a paired \textit{t}-test was performed to identify significant connectivity changes in Broca and Wernicke's area during imagined speech and resting states.

\section{Results and Discussion}

\subsection{Decoding Performance}
We developed the frameworks of decoding the speech-related EEG signals of 13 classes in two conditions of imagined speech and overt speech. 
The performance of imagined speech and overt speech was compared. The average accuracy of overt speech was 49.5\% for 13 classes, including 9 subjects' performances. The EEG signal during overt speech can contain more significant representation in brain activities. As we conducted preprocessing to remove artifacts related to EOG and EMG around mouth, the EEG signal only contains the brain activity to intent to move mouth and tongue to speak out the pronunciation for each word. The average accuracy of imagined speech was 35.07\% for 13 classes, including 9 subjects' performances. The EEG signal during imagined speech includes only brain activities rather than EMG since they did not move their muscle. Therefore, the performance of imagined speech normally inferior than it of overt speech. However, the difference between overt speech and imagined speech was significantly different($p<0.05$), but not so huge while overt speech was expected to show superior performance.

\subsection{Attention Module}
We showed that deep learning model with self-attention module could show reasonable performance. The advantages of the self-Attention module are that it can reduce the total computational complexity per layer, that it can parallelize the computational volume to some extent, and that the path length of long-term dependency in the networks is short.

%The attention module had several parameters that can impact to the performance such as embedding size, the number of attention heads, hidden layer size, and etc. The parameters of attention module can be optimized for improving the performances. 

\section{Conclusion}
In this study, we proposed attention module based on transformer architecture to decode imagined speech in EEG. As practical BCIs require a robust system and simple hardware usable in the real-world, we show that the proposed method improved the BCI performance.
The results of recognizing speech from human intention had reasonable performance although we used only few channels. And we compared overt speech and imagined speech in terms of performance and statistical analysis. The EEG of overt speech showed superior performance than imagined speech, which was significant different, but not that huge than we expected. Therefore, technology of decoding imagined speech with attention module had potential to use as a real-world communication system.
In the future, we developed the architecture that performed with higher performance for imagined speech. Moreover, parameter optimization of self-attention module can increase the performance as well.

\bibliographystyle{IEEEtran}
\bibliography{mybib}

\end{document}